\documentclass[9pt]{article}
\usepackage{spconf}
\usepackage{flushend,cite}
\usepackage{amsmath,amssymb,amsfonts}
\usepackage{algorithmic}
\usepackage{graphicx}
\usepackage{textcomp}
\usepackage{xcolor}
\def\BibTeX{{\rm B\kern-.05em{\sc i\kern-.025em b}\kern-.08em
    T\kern-.1667em\lower.7ex\hbox{E}\kern-.125emX}}
\usepackage[hyphens]{url}
\usepackage{hyperref}
\usepackage{booktabs}
\usepackage{tikz}
\usetikzlibrary{positioning}
\usetikzlibrary{shapes.geometric}

\name{\begin{tabular}{c}Francesco Mauro$^{a,1}$,  Benjamin Rich$^{b}$, Veronica Wairimu Muriga $^{b}$,\\
\textit{Alessandro Sebastianelli}$^{c}$ \textit{and Silvia Liberata Ullo}$^{a}$\end{tabular}\thanks{
$^{1}$Corresponding author. 
\textit{Email addresses}: f.mauro$@$studenti.unisannio.it (FM), brrich$@$mit.edu (BR), wmuriga$@$mit.edu (VWM), alessandro.sebastianelli$@$esa.int (AS), ullo$@$unisannio.it (SLU)}
}

\address{
$^{a}$ Engineering Department, University of Sannio, Benevento, Italy \\
$^{b}$ Massachusetts Institute of Technology, Boston, USA \\
$^{c}$ $\phi$-lab, European Space Agency, Frascati, Italy\\
}

\title{Sen2DWater: a novel multispectral and multitemporal Dataset and Deep Learning Benchmark for Water Resources Analysis}

\begin{document}
\maketitle

\begin{abstract}
Climate change has caused disruption in certain weather patterns, leading to extreme weather events like flooding and drought in different parts of the world. 
In this paper, we propose machine learning methods for analyzing changes in water resources over a time period of six years, by focusing on lakes and rivers in Italy and Spain. Additionally, we release open-access code to enable the expansion of the study to any region of the world.
 We create a novel multispectral and multitem\nobreak poral dataset, SEN2DWATER, which is freely accessible on GitHub. We introduce suitable indices to monitor changes in water resources, and benchmark the new dataset on three different deep learning frameworks: Convolutional Long Short Term Memory (ConvLSTM), Bidirectional ConvLSTM, and Time Distributed Convolutional Neural Networks (TD-CNNs). 
Future work exploring the many potential applications of this research is also discussed.
\end{abstract}
\begin{keywords}
Climate change, Water Resources, Drought, Sentinel-2, Water Indices, Deep Learning
\end{keywords}
\vspace{-0.3cm}
\section{Introduction}
\label{sec:intro}
\vspace{-0.3cm}
Climate change has had significant negative environmental effects on our planet, leading to disruption of weather patterns and causing unusual and extreme weather (\href{https://www.unicef.org/stories/water-and-climate-change-10-things-you-should-know}{Water and the global climate crisis: 10 things you should know}). Abnormal weather causes flooding in some parts of the world, while other regions suffer from drought and water scarcity. Organizations worldwide have highlighted the importance of taking necessary and timely measures to curb the effects of climate change before they become irreversible. The United Nations has established 17 Sustainable Development Goals (SDGs) in the 2030 Agenda for Sustainable Development (\href{https://sdgs.un.org/goals}{Sustainable Development Goal}) and the availability of water resources plays a crucial role in meeting many of these goals. We propose research for predicting changes in water availability, by using data collected over the last six years. We propose neural network architectures for analyzing changes in water resources, and train on data collected through remote sensing methods. We make available open-access code to enable the expansion of the study to other geographic areas. We take inspiration from (\href{https://global-surface-water.appspot.com/#}{Global Surface Water Explorer}), which maps the location and temporal distribution of water surfaces at a global level using Landsat data.

By collecting Sentinel-2 data from July 2016 to December 2022, we create a new dataset, SEN2DWATER, which to the best of our knowledge, is unique in several aspects with respect to other already available datasets.(\href{https://www.kaggle.com/datasets/mateuszst/water-body-segmentation-from-satellite-images}{Water Body Segmentation From Satellite Images}, \href{https://www.kaggle.com/datasets/franciscoescobar/satellite-images-of-water-bodies}{Satellite Images of Water Bodies}, \cite{sui2022high}, \cite{dat2016global}, \cite{p2016high}). SEN2DWATER is a 2 Dimensional (2D) Spatiotemporal dataset using data from all 13 bands of Sentinel-2, with high spatial resolution compared to other related remote sensing satellites. SEN2DWATER contains data collected over a six-year period, using remote sensing methods, of water basins in Italy and Spain. Our code is accessible online to allow for the extension of the dataset.

Table 1 summarizes the differences between our dataset and other datasets specifically created for the purposes of water resources analysis.
\vspace{-0.45cm}
\begin{table}[!ht]
    \centering
    \caption{Categorization of SOTA datasets} \label{tab:categorization}
    \resizebox{1\columnwidth}{!}{
    \begin{tabular}{llllll}
        \toprule
        Paper/Dataset & Satellite & Multispectral & Multitemporal & Resolution & $N^\circ$ Samples\\
        \toprule
        \href{https://www.kaggle.com/datasets/franciscoescobar/satellite-images-of-water-bodies}{Satellite Images of Water Bodies} & Sentinel-2 & No & No & 10m & 2841 \\
        \href{https://www.kaggle.com/datasets/mateuszst/water-body-segmentation-from-satellite-images}{Water Body Segmentation From Satellite Images} & Sentinel-2 & No & No & 10m & 10 \\
        \cite{sui2022high} & Sentinel-2 & - & No & 10m & - \\
        \cite{dat2016global} & Landsat-7 ETM+ & Yes & No & 30m & 8756\\
        \cite{p2016high} & Landsat & No & Yes & 30m & 3000000 \\
        \midrule
        SEN2DWATER & Sentinel-2 & Yes & Yes & 10m & $39\times5264=205296$ \\
        \bottomrule
    \end{tabular}}
\end{table}

\begin{figure*}[!ht]
    \centering
    \includegraphics[width=1.6\columnwidth]{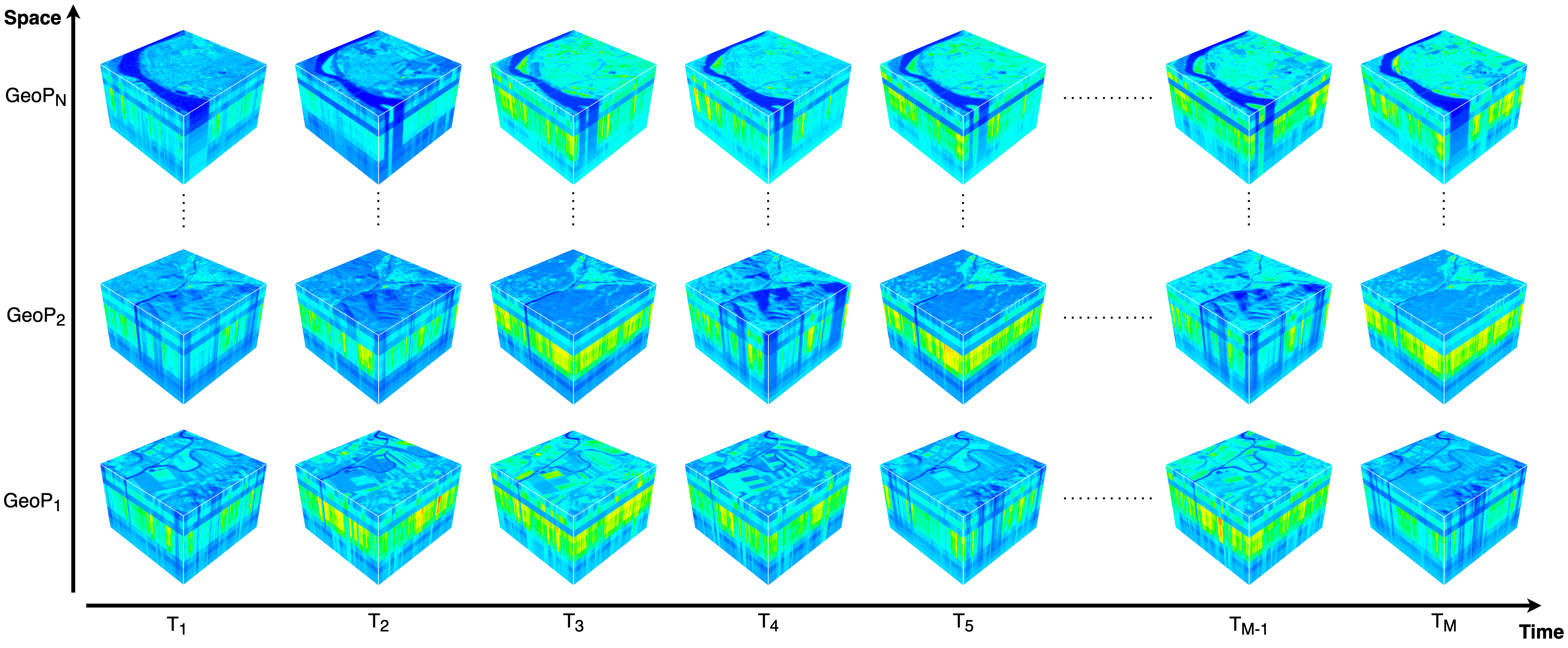}
    \caption{Visualization of the SEN2DWATER Dataset. On the y-axis are different geographical locations, and on the x-axis are the time series images for each location. Each cube plot shows the Sentinel-2 product composed of 13 spectral bands.}
    \label{fig:dataset}
\end{figure*}

We perform a state-of-the-art study by searching for the most suitable neural network capable of taking into account both spatial and temporal features, such as those present in SEN2DWATER, to yield desirable and useful results relating to water resources analysis. 
In this study, we benchmark our dataset on the three following Deep Learning frameworks: $1)$ Convolutional Long Short-Term Memory (ConvLSTM), $2)$ Bidirectional ConvLSTM and $3)$ Time Distributed Convolutional Neural Network (CNN) \cite{shi2015convolutional, liu2017bidirectional, dutt2021automatic}.

Using SEN2DWATER, suitable indicators for this analysis can be extracted, including Normalized Difference Water Index (NDWI), Modified NDWI (MNDWI), Weighted NDWI (WNDWI), and Normalized Difference Vegetation Index (NDVI).
The dataset and code used to extract these indices can be found on our GitHub page.

This research has several potential avenues for further exploration: extending the dataset and related analysis to other geographic areas, longer periods of analysis, use of specific neural networks for water shrinkage forecasting, the inclusion of water-specific indicators, availability of multispectral data for other use cases besides water analysis, etc. We believe that this research area has great potential to help combat climate change. 
\vspace{-0.5cm}
\section{Dataset definition and creation} \label{sec:sent2}
\vspace{-0.25cm}
The use of machine learning (ML) techniques is growing in the field of remote sensing, with applications ranging from detection and classification of land cover and use to the monitoring and prediction of numerous natural or anthropogenic hazard events. A major barrier to the practical use of these machine learning techniques for remote sensing applications 
is the extensive amount of data required for training the particular neural network, as well as the ensuing computing load, and the time required to collect the data \cite{sebastianelli2021automatic}. 
By creating our dataset, SEN2DWATER, we hope to reduce barriers involved with remote sensing research by making available an expansive, geographically diverse dataset that can be used as-is or readily adapted for other use cases.  

In our case, to evaluate the water resources among the specific area of interest, we create a novel dataset by downloading Sentinel-2 (Sen2) data. The Sen2  mission consists of two polar-orbiting satellites with a large swath width (290 km) and a high revisit time (10 days at the equator with one satellite and 5 days with two satellites in cloud-free circumstances). Each satellite hosts a Multi-Spectral Instrument (MSI) measuring the Earth's reflected radiance in 13 spectral bands at different spatial resolutions: four bands at $10 m$, six at $20 m$, and three at $60 m$ (\href{https://global-surface-water.appspot.com/#}{MultiSpectral Instrument (MSI) Overview}),(\href{https://sentinel.esa.int/web/sentinel/missions/sentinel-2}{ESA: Sentinel-2}).

The built dataset is shown in Fig. \ref{fig:dataset}, where $N$ represents the number of different geographical points $(GeoP_n)$ and $M$ the length of the time series. Therefore, all time series are of equal length. Namely, Sen2A and Sen2B multispectral data related to selected water basins has been downloaded from the \textit{Google Earth Engine} (GEE) platform  from July 2020 to December 2022. The less cloudy sample within a two-month time span has also been automatically identified and selected as part of the dataset. Each image $GeoP_n^{(m)}$ contains all 13 bands of Sen2, where the bands at resolution higher than $10 m$ have been re-sampled to $10 m$. The pipeline used to download the dataset is a modified version of the one proposed in \cite{sebastianelli2021automatic}. 
It is worth highlighting that after 2022-01-25, Sen22 scenes with PROCESSING\_BASELINE "04.00" or above had their DN range shifted by 1000 (\href{https://developers.google.com/earth-engine/datasets/catalog/COPERNICUS_S2_HARMONIZED}{Harmonized Sentinel-2 MSI: MultiSpectral Instrument, Level-1C}). Therefore, we switch to using the HARMONIZED collection in GEE, which takes into account the shift in range.

\begin{figure}
    \centering
    \resizebox{0.7\columnwidth}{!}{
    \begin{tabular}{cc}
        \includegraphics[width=0.7\columnwidth]{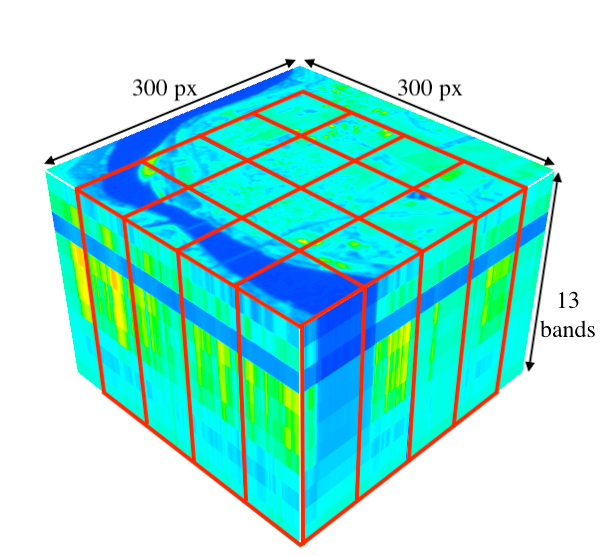} & 
        \includegraphics[width=0.3\columnwidth]{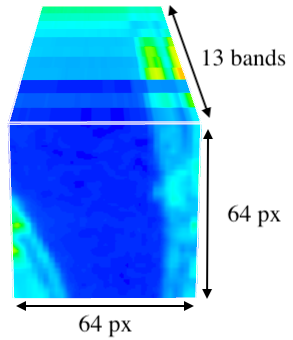}\\ 
    \end{tabular}}
    \caption{Process of patches extraction.}
    \label{fig:patches}
\end{figure}

Since each downloaded image is from a geographical area of $3km\times 3km$ with a spatial resolution of $10 m$,  each image is made up of $300px\times 300px$. Therefore, the dataset $D_1$ is defined in $D_1 \in R ^{(Geo \times Time \times Width \times Height \times Spectrum)}$, where in our case $Geo=329$, $Time=39$, $Width=300$, $Height=300$ and $Spectrum=13$. In order to increase the size of the dataset we extracted from each image of $300px\times 300px$ $16$ patches of $64px\times 64px$ (discarding the part of the images that remain since 300 is not divisible by 64) as shown in Fig. \ref{fig:patches}.  The image is divided into 16 patches and is not further divided as decreasing the width and height of the image past this point makes it difficult to clearly identify the characteristics of the image (profile of a water basin, land, etc.). The new dataset $D_2$ is defined by $D_2 \in R ^{(16 \times Geo \times Time \times Width \times Height \times Spectrum)}$, where in our case $Geo=5264$, $Time=39$, $Width=64$, $Height=64$ and $Spectrum=13$.
\section{Definition of Indices} \label{sec:indexes}
\vspace{-0.35cm}
 There are many possible applications of the SEN2DWATER dataset. It is possible to differentiate between water and other surface materials by exploiting the spectral reflectance properties of water. Changes in water resources can be monitored by analyzing the value of suitable indices \cite{2022surface}.
The presence of water can be determined through the NDWI (\ref{NDWI1_Formula}), used to track changes in water bodies. The near-infrared (NIR) and green electromagnetic spectrum bands are used to identify water bodies because water heavily absorbs light in the visible to infrared electromagnetic range. NDWI is sensitive to built-up land and can result in over-estimation of water bodies. NDWI was proposed by McFeeters in 1996 (\href{https://custom-scripts.sentinel-hub.com/custom-scripts/sentinel-2/ndwi/}{NDWI Normalized Difference Water Index}) and has a range of values between -1 and 1. 
Positive values of NDWI indicate the presence of water\cite{2022surface} \cite{guo2017weighted}.   
\begin{equation}
    NDWI = \frac{Green - NIR}{Green + NIR}
    \label{NDWI1_Formula}
\end{equation}

The NDWI was modified by substituting the NIR band with the short-wave infrared band (SWIR) \cite{MDPI2016water}. The MNDWI can suppress built-up land noise as well as vegetation and soil noise, avoiding the overestimation of the presence of water. Positive values of MNDWI again indicate the presence of water. \cite{xu2006modification}. 
However, using NDWI to classify turbid water can lead to incorrect results, and using MNDWI could lead to incorrectly classifying the shadowy areas of vegetation as water \cite{guo2017weighted}. To reduce these errors, \cite{guo2017weighted} proposes a weighted index, the WNDWI. In (\ref{wndwiForm}), $a \in [0, 1]$ is a weighted coefficient. When $a = 1$, the WNDWI is equivalent to
NDWI and when $a = 0$, the WNDWI is equivalent to MNDWI.
\begin{equation}
    WNDWI = \frac{Green - a \cdot NIR - (1-a) \cdot SWIR}{Green + a \cdot NIR + (1-a) \cdot SWIR}
    \label{wndwiForm}
\end{equation}
Lastly, another index, the NDVI (\ref{NDVI_Form}), is often used for a complete analysis of water resources. Negative NDVI values indicate the presence of water \cite{2022surface}.\
\begin{equation}
    NDVI = \frac{NIR - Red}{NIR + Red}
    \label{NDVI_Form}
\end{equation} 
Note that for the Sen2 dataset,  Green, Red, NIR and SWIR are respectively B3, B4. B8 and B11.
\vspace{-0.35cm}
\section{DEEP LEARNING  BENCHMARK}\label{sec:dataset_benchmarking}
\vspace{-0.22cm}
Since the proposed dataset contains time-series data, we can benchmark our dataset through a next-frame prediction task. This set of experiments is a starting point for benchmarking our dataset and will be extended in future work.

We selected three deep learning frameworks able to handle time series of images: $1)$ a ConvLSTM model, $2)$ a Bidirectional ConvLSTM model and a $3)$ Time Distributed Convolutional Neural Network. We use these three frameworks for next-frame prediction, where a new image is predicted using a time series of past acquisitions. In our case, we predict the NDWI map 2 months ahead using the last 7 acquisitions. Our \href{https://github.com/francescomauro1998/Impact-of-Climate-Change-on-Water-Resources}{code} is modular, and therefore it is easy to use for the testing of different networks with different combinations of parameters, including the length of the time series and the predicted spectral index. Our \href{https://github.com/francescomauro1998/Impact-of-Climate-Change-on-Water-Resources}{GitHub page} contains the TensorBoard records for the models used in this paper in order to make our work fully reproducible- to conserve memory we released only the final simulation for each model. These records contain the model weights and the numerical and visual results obtained after training.

\begin{figure}[!ht]
    \centering
    \includegraphics[width=0.75\columnwidth]{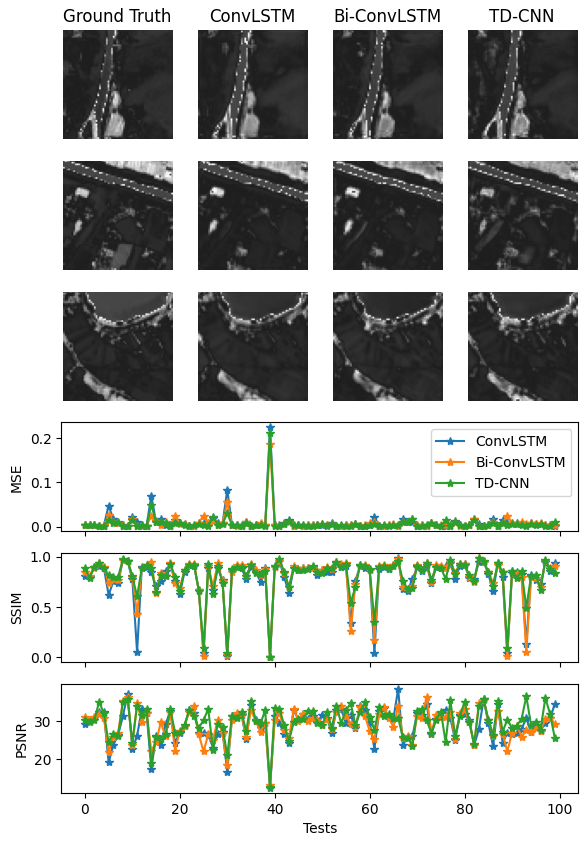}
    \caption{Results for the proposed models on validation set.}
    \label{fig:results}
\end{figure}

Results are reported in Fig.\ref{fig:results}, which is split into two parts. Each row of the top grid shows a visual result for the proposed models on the validation set when compared with the Ground Truth (GT). The bottom part shows the trends of Mean Squared Error (MSE), Structural Similarity Index (SSIM), and Peak Signal to Noise Ratio (PSNR) for 100 samples of the validation set. Averaged numerical results on the validation set are reported in Table \ref{tab:results}.

\begin{table}[!ht]
    \centering
    \resizebox{0.8\columnwidth}{!}{
    \begin{tabular}{lccc}
    \toprule
    Score            & ConvLSTM & Bi-ConvLSTM & TD-CNN \\
    \midrule
    MSE $\downarrow$ & $ 0.010 \pm 0.025$ & $ 0.008 \pm 0.019$ & $ 0.008 \pm 0.022$\\
    SSIM $\uparrow$  & $ 0.784 \pm 0.225$ & $ 0.799 \pm 0.220$ & $ 0.810 \pm 0.188$\\
    PSNR $\uparrow$  & $29.066 \pm 4.255$ & $29.214 \pm 3.901$ & $29.983 \pm 3.974$\\
    \bottomrule
    \end{tabular}}
    \caption{Averaged numerical results on validation set.}\label{tab:results}
\end{table}
Visual and numerical results show that TD-CNN performs better than ConvLSTM and Bi-ConvLSTM with respect to MSE, SSIM and PSNR evaluation metrics. However, the difference in the performance of these three frameworks is not large. The Bi-ConvLSTM presents less variability with respect to the TD-CNN. Qualitatively, each framework produces visually accurate images relative to the ground truth. However, TD-CNN preserves more detail relative to the other frameworks. We find these preliminary results to be very encouraging  as they allow us to predict changes in a geographic area with limited error. 
\vspace{-0.55cm}
\section{CONCLUSIONS}
\vspace{-0.25cm}
In this paper, we present a novel dataset that can be used to analyze water basins. We leverage three deep learning frameworks and exploit a specific index, NDWI, to predict changes in water basins. We present promising preliminary results which demonstrate that NDWI can be predicted in advance with a limited deviation from the GT. A more intensive hyperparameter search will be implemented in future work to identify a more robust and better-performing solution. This research has several potential avenues for further exploration, and this work, in addition to future work, has the potential to help in combating climate change. We are expanding SEN2DWATER to include a wider spatial distribution and are further benchmarking the dataset. 

\vspace{-0.35cm}
\section{Acknowledgments}
\vspace{-0.22cm}
\noindent This research work began as the Master's Thesis of Francesco Mauro, jointly supervised by Alessandro Sebastianelli and Silvia Liberata Ullo, and will be further developed during his Ph.D. The paper is part of a joint project between the University of Sannio and the Massachusetts Institute of Technology (MIT), involving students Veronica Muriga and Benjamin Rich. 

\vspace{-0.45cm}
\bibliographystyle{IEEEbib}
\bibliography{refs}

\end{document}